\documentclass[12pt]{iopart}

\usepackage{iopams} 
\usepackage{hyperref}
\usepackage[capitalize]{cleveref}
\usepackage{siunitx}
\usepackage{lscape}
\usepackage{xcolor}
\usepackage{graphicx}
\usepackage{placeins}
\usepackage{float}
\usepackage{subfigure}
\usepackage{graphicx}
\usepackage{natbib}
\newcommand{\au}{\,\mathrm{au}}
\begin{document}

\title[Simulator for the near-Sun thermal environment]{SHINeS: Space and High-Irradiance Near-Sun Simulator}

\author{Georgios Tsirvoulis$^1$, Mikael Granvik$^{1,2}$ and 
Athanasia Toliou$^1$}

\address{$^1$ Asteroid Engineering Laboratory, Space Systems, Lule\r{a} University of Technology, Box 848, SE-98128 Kiruna, Sweden}
\address{$^2$ Department of Physics, PO Box 64, 00014 University of Helsinki, Finland}
\ead{georgios.tsirvoulis@ltu.se}

\vspace{10pt}
\date{}

\begin{abstract}
We present SHINeS, a space simulator which can be used to replicate the thermal environment in the immediate neighborhood of the Sun down to a heliocentric distance$\mathrm{r}\sim0.06\au$. The system consists of three main parts: the solar simulator which was designed and constructed in-house, a vacuum chamber, and the probing and recording equipment needed to monitor the experimental procedures. Our motivation for building this experimental system was to study the effect of intense solar radiation on the surfaces of asteroids when their perihelion distances become smaller than the semi-major axis of the orbit of Mercury. Comparisons between observational data and recent orbit and size-frequency models of the population of near-Earth asteroids suggest that asteroids are super-catastrophically destroyed when they approach the Sun. Whereas the current models are agnostic about the disruption mechanism, SHINeS was developed to study the mechanism or mechanisms responsible. The system can, however, be used for other applications that need to study the effects of high solar radiation on other natural or artificial objects.
\end{abstract}

%
%
%
\maketitle
%
%

\section{Introduction}
In recent years, owing to the advancements in telescopic surveys of the night sky in search for new asteroids, a new class of small solar system bodies has been established; that of active asteroids \citep{aste2015}. These are objects that have orbital characteristics of asteroids, but they also feature activity in the form of mass loss similar to comets. Many different mechanisms have been proposed to explain their activity, including sublimation of water ice, impacts with smaller bodies, rotational fission, electrostatic levitation of particles, radiation pressure sweeping, and thermal effects. 

Of particular interest to us are the ones that are also categorized as near-Earth asteroids (NEAs), i.e., having perihelion distances $q<1.3\au$. Their small $q$ means that, having spent enough time relatively close to the Sun, they should be devoid of any water ice that could drive their activity. The prime example of such an asteroid is (3200)~Phaethon, which is responsible for the annual Geminid meteor stream due to its generated dust tail. However, the exact mechanism driving its activity is not fully understood. Thermal effects such as desiccation cracking, that is the formation of cracks and ejection of material due to the rapid drying of its hydrated minerals during perihelion passages, are the most probable explanations \citep{Jewitt2013}.

A recent development in understanding the activity of NEAs was due to \citet{Granvik2016} during the development of their state-of-the-art model of the near-Earth-object population. The number of asteroids with small $q$ that their model predicted that should have been observed was significantly larger than what was actually observed. Therefore, they came to the realisation that there has to exist a mechanism by which asteroids are super-catastrophically destroyed when their orbits evolve close to the Sun. The best-fit estimate for the disruption distance, when averaging over all the relevant parameters such as the sizes, orbital parameters and geometric albedos, was computed at $q=0.076\au$. Indeed, without taking into account the proposed mechanism, the population model over-predicts the total number of NEAs by around 5\%, whereas for small $q$ the over-prediction grows to 900\%. They also showed that the disruption distance and the asteroid sizes are correlated, as smaller asteroids tend to break up easier, therefore farther away from the Sun compared to larger ones. 

Another prediction of their model, coming from the extrapolation of the estimated disruption distance for asteroids of 40 meters in diameter to that of 1 meter objects, was that such small objects should not exist interior to the orbit of Mercury. \citet{Brown2016} verified this prediction by studying the orbits of the parent bodies of 59 recorded airbursts, and concluded that there are no Earth impactors with diameters smaller than one meter and $q$ inside the orbit of Mercury. \citet{Wiegert2020} also verified this prediction based on data of NEAs and SOHO comets, and they concluded that entire asteroids with diameters on the order of kilometers can be destroyed into millimeter sized fragments through a cascade of disruption events.

In the past there have been some studies aiming to examine the effects of heating on asteroids, but with completely different scope and methods. \citet{Delbo2014} studied the generation of regolith due to crack growth induced by thermal cycling, but this was done for a completely different temperature range (corresponding to distances from the Sun of $1-2.5\au$) by using electrical elements as heat sources, under atmospheric pressure. 
\citet{masiero2021} used heating elements to heat samples of meteorite Allende in a temperature range of $300-\ang{800}\mathrm{C}$ in a vacuum chamber,  to study the volatility of Sodium as the primary mechanism responsible for Phaethon's activity. On the other hand, \citet{Dreyer2016,Gertsch2017} performed experiments to assess the viability of optical mining technology using simulated or concentrated solar light on minerals, using much higher power fluxes than the ones relevant in the solar neighborhood. 

We decided to take a different approach for our experiments, one that directly simulates the solar irradiation at distances down to at least $0.076\au$ on asteroid-like minerals, rather than trying to achieve modelled or measured equilibrium temperatures by using heating elements. As such, we present here the design, development and first tests of SHINeS: Space and High-Irradiance Near-Sun Simulator, which is based on a vacuum chamber and a high-irradiance light source, as well as a number of measuring and monitoring sub-systems.

In Section 2, we demonstrate the design process and the details of the system's hardware. In Section 3, we present the necessary calibration phases regarding the uniformity, optical power and spectral characteristics of the solar simulator. In Section 4, we present a case study of an experiment which we ran to verify the functionality of all sub-systems. Finally, we present some conclusions of our work in Section 5. 

\section{Experimental setup}

SHINeS is comprised of three main parts: a) a vacuum chamber, b) a solar simulator and c) the required measuring and monitoring devices, which are presented below.

\subsection{Vacuum chamber}

\begin{figure}[h]
\centering
\includegraphics[width=0.77\textwidth]{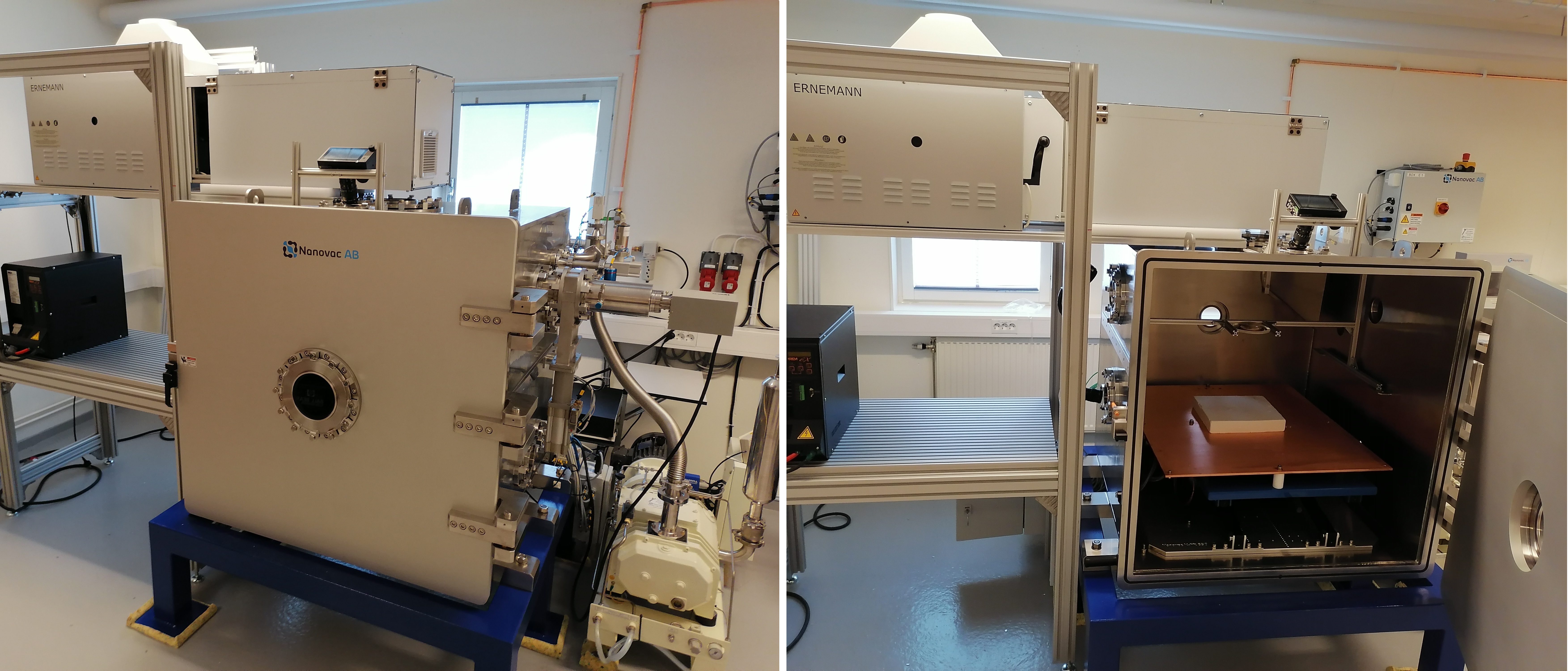}
\caption{The vacuum chamber and the main components.}
\label{fig:chamber}
\end{figure}

The vacuum chamber was manufactured by Nanovac AB, Sweden, designed to meet the unique requirements of our intended experiments. Since our primary use will be to irradiate mineral samples in it, which are expected to result in considerable outgassing, the size was decided to be relatively large in order to maintain as low pressure as possible. The result is a $1\,\text{m}\times1\,\text{m}\times1\,\text{m}$ stainless steel chamber of cubical shape (\cref{fig:chamber}). The chamber's various functions can be operated either locally through the Human-Machine Interface (HMI) or remotely via a VPN connection.

For pumping down, the chamber features a Hanbell PSE-80 dry screw  roughing pump (water cooled via a SMC Co. circulation chiller) and a Leybold MagIntegra 1300 iD turbomolecular high vacuum pump. The system has the capability to achieve a minimum pressure of $<10^{-7}\,$mbar. The pumping speed depends on a number of factors, with ambient air humidity being the largest one, but it is generally able to achieve a pressure of $10^{-4}\,$mbar (the operational threshold of the RGA as will be discussed below) within a few hours, and a pressure of $10^{-5}\,$mbar (our practically chosen point for starting the experiments) usually achieved in less than 24 hours, allowing for a two-day workflow (setup on the first day, experiment execution on the following day). Since our experiments will feature mineral samples which might fragment under the radiation, the port of the turbo pump has to be protected from potential debris. We installed a stainless steel shield that blocks the port's line of sight view towards the center of the chamber where the samples will be placed, as well as a $\SI{45}{\um}$ stainless steel mesh to filter out potential incoming dust particles.

\begin{figure*}[h]
\centering
\includegraphics[width=\textwidth]{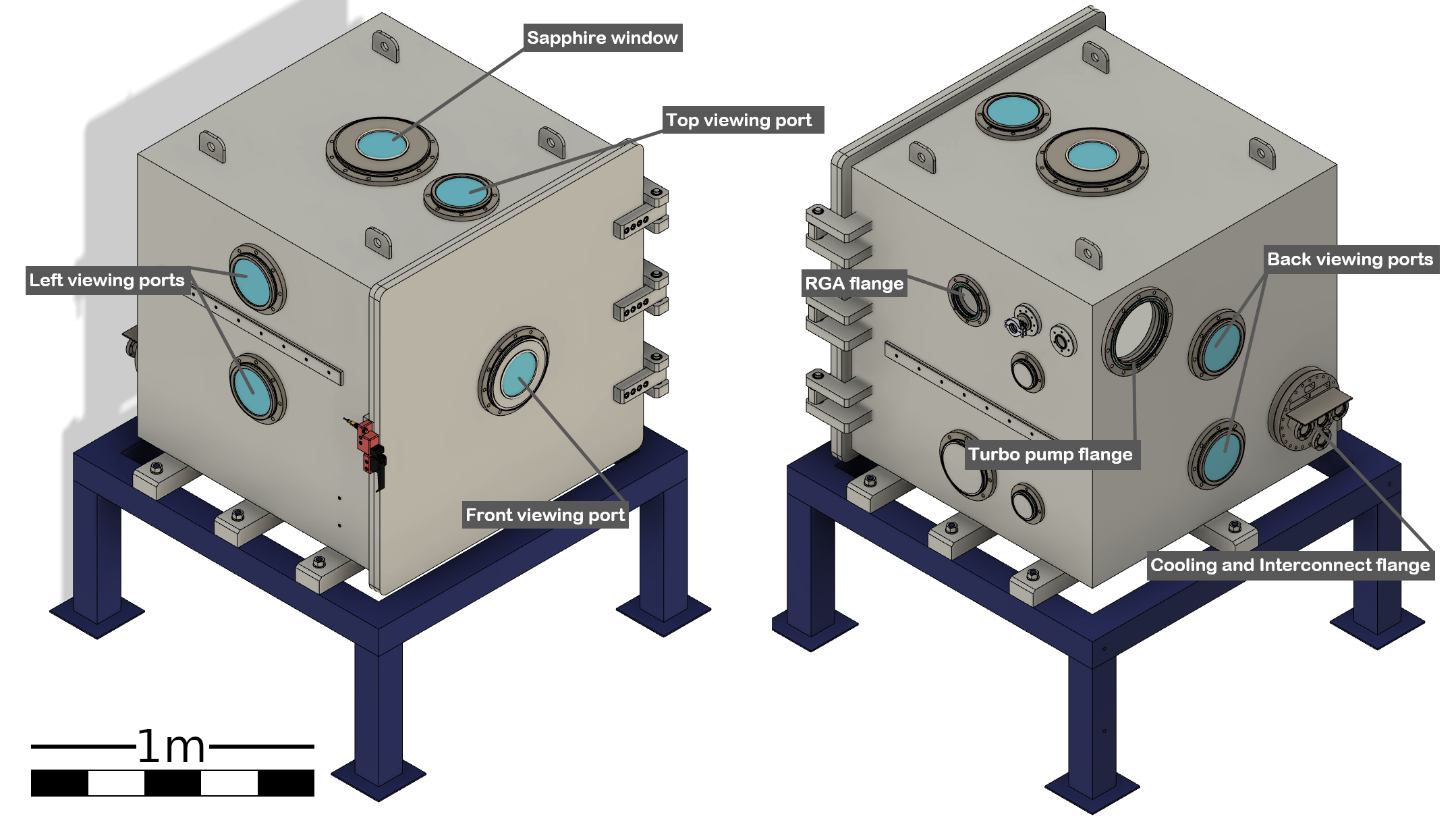}
\caption{CAD model of the vacuum chamber showing the location of the viewing and instrumentation port locations.}
\label{fig:cad}
\end{figure*}

At the center of the top face of the vacuum chamber there is a 130-mm-diameter window through which the light beam enters the chamber. The window is made of sapphire glass due to its flat spectral  transmittance and good thermal performance. Particular emphasis was put on the ability to monitor the processes happening inside the chamber during the experiments, therefore, the design facilitates a number of viewing ports on all the faces of the vacuum chamber apart from the bottom one (\cref{fig:cad}). The viewing port located on the top face allows for a clear top-down view of the sample. The top windows on the left and back faces of the chamber give a wide $\sim\ang{45}$ view while the bottom ones give an edge on side view.

On the right face of the chamber, there is a flange where a secondary cylindrical volume is connected, which houses  a residual gas analyser (RGA). The secondary volume can be isolated from the main chamber volume via a valve. More details on the RGA will be presented in the following subsections.

The interior of the vacuum chamber is equipped with a copper table plate, which is adjustable in height by means of a manually-operated scissor-type lifting mechanism. The table has the capability of serving as a chiller plate as it has a copper tube welded on its underside, specified to work with liquid Nitrogen. This was implemented to accommodate for experiments that would require cycles of heating and cooling of the samples. Several mounting points have been machined on the copper table for securing various equipment or sample holders.

For connecting different instruments that need to be inside the chamber there are two general purpose D-sub 25 as well as two USB 3.0 feedthrough connectors. Additional blind ports are also available in case more feedthrough connectors need to be installed for future projects.

\subsection{Solar simulator}

The experiments we are planning on conducting require some special characteristics for the irradiation of the samples. The light source needs to be powerful enough and to be concentrated at the sample position inside the vacuum chamber, in order to deliver the necessary incident optical power flux. This fact, combined with the large dimensions of the chamber itself, deemed it impossible to find an off-the-shelf solar simulator solution that would satisfy our needs. Therefore, we designed and built the light source  apparatus ourselves. The system we developed is comprised of three main parts, namely  the lamp housing, the power supply, and the optical path.

\subsection*{Light source}

The three most common types of lamps used for solar simulator designs are Xenon arc, metal Halide and Argon arc. For a detailed review on the characteristics, advantages and disadvantages, the reader is referred to \citet{Gallo2017}. For our design we selected a Xenon arc source, as its common use in cinema-theater projectors made it possible to acquire commercial-off-the-shelf components such as the lamp housing, power supply and the lamps themselves, while maintaining good spectral characteristics and adequate optical power output. Another key advantage is that the spectral profile is independent of the input electrical power level of operation, or small fluctuations of it, which will allow us to run our experiments at various irradiation levels.

A Xenosol 7000 lamp housing, typically used in the Ernemann 15 cinema theater projector, was the solution that met our power and output light beam requirements. As the name suggests, it can be used with Xenon-arc lamps of up to 7~kW of power, and it contains the necessary electronic sub-assembly for igniting the lamp as well as two high-flow filtered blowers to cool the lamp. It features an ellipsoidal-by-revolution first-surface mirror, with a focal length of $700\,$mm. The mirror reflects $\sim$61\% of the light generated by the arc of the lamp, which is placed at its primary focus, toward the secondary focus, which is located $\sim21\,$cm after the aperture of the housing. 

The lamp we use in the current setup is an OSRAM XBO 7000 W/HS XL OFR, an industry standard for Xenon arc lamps, with a nominal maximum electrical power of 7000~W as the name suggests.  

\subsection*{Power supply}

The power supply plays an important role in the design of our experimental setup, with the two primary factors in our selection being the wide power range of operation and the stability of the output electrical power. Our selection was the EX-170GM3-E, manufactured by IREM, Italy. Its internal electronic controls ensure stable power delivery in a range of operation of 3--7~kW, which can be easily adjusted  to the desired level for each application as will be discussed in the following. 

\subsection*{Beam formation}

Since the sapphire window of the vacuum chamber, through which the light enters, is located at the center of its top face, there is a considerable distance between the secondary focus of the lamp housing and the chamber's work table, where the samples will be placed. As the light exits horizontally from the lamp housing, the beam needs to travel approximately $700\,$mm horizontally, then be reflected downwards into the chamber's window, and be concentrated onto the sample. In order to achieve this, the initial design idea was the following. After the secondary focus of the ellipsoidal mirror the light forms a diverging cone. We would make use of a convex lens to parallelize it and transmit it horizontally roughly the whole distance to the sapphire window. Then a second, long focal length, convex lens would be placed to focus the parallel beam at its focus. A flat mirror at a $\ang{45}$ angle would then be placed above the sapphire window to turn the converging beam downwards into the vacuum chamber and onto the test table. Unfortunately, reality is once again more challenging than the ideal case. 

The biggest factor working against us was the fact that the light-emitting arc inside the lamp is not a point source, but rather an extended source, albeit a small one. Since the first parallelizing lens of our design has a short focal length, and the second, the re-focusing one, has a much longer focal length, the image of the extended source is magnified at the focal plane which is the test table. Added to that, although with a much smaller effect, are the spherical aberrations induced by the lenses themselves, resulting in a circular illuminated area with a diameter of approximately 300mm on the focal plane, and a low luminous power density. To overcome these challenges we modified our design, mainly by including a second focusing lens, placed inside the vacuum chamber. This lens brings the focal plane closer to the source, but concentrates the light to a much smaller spot, resulting in a much higher optical power density.

\begin{figure}[h]
\centering
\includegraphics[width=0.6\textwidth]{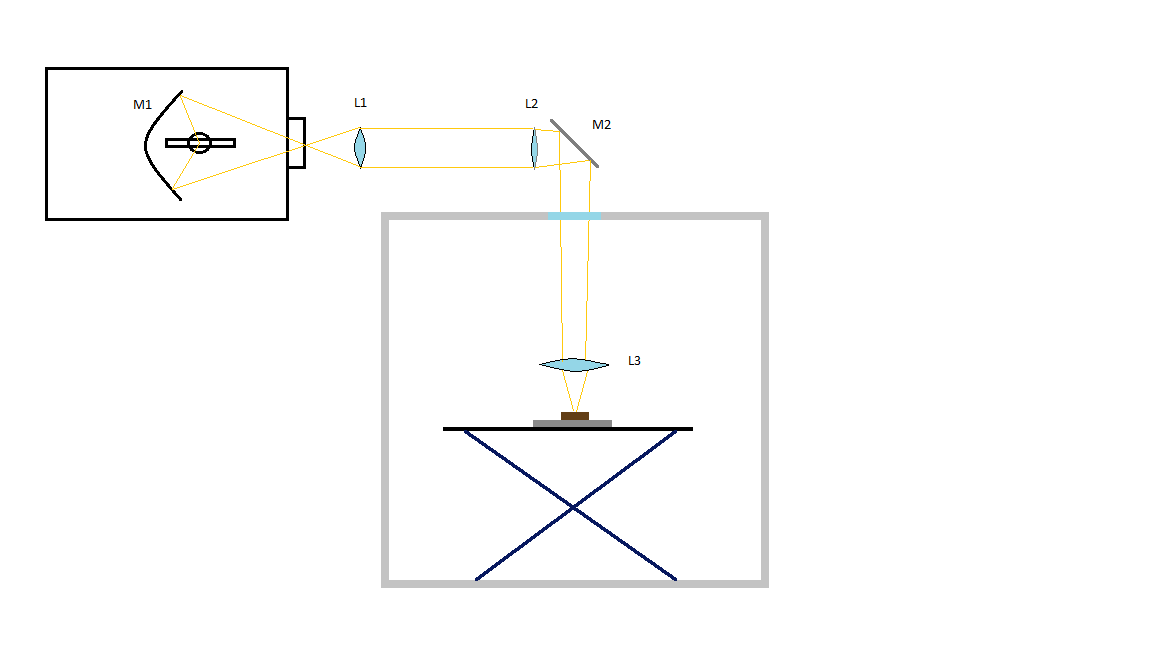}
\caption{Sketch of the optical path assembly (not to scale).}
\label{fig:lightsource}
\end{figure}

\begin{figure}[h]
\centering
\includegraphics[width=0.47\textwidth]{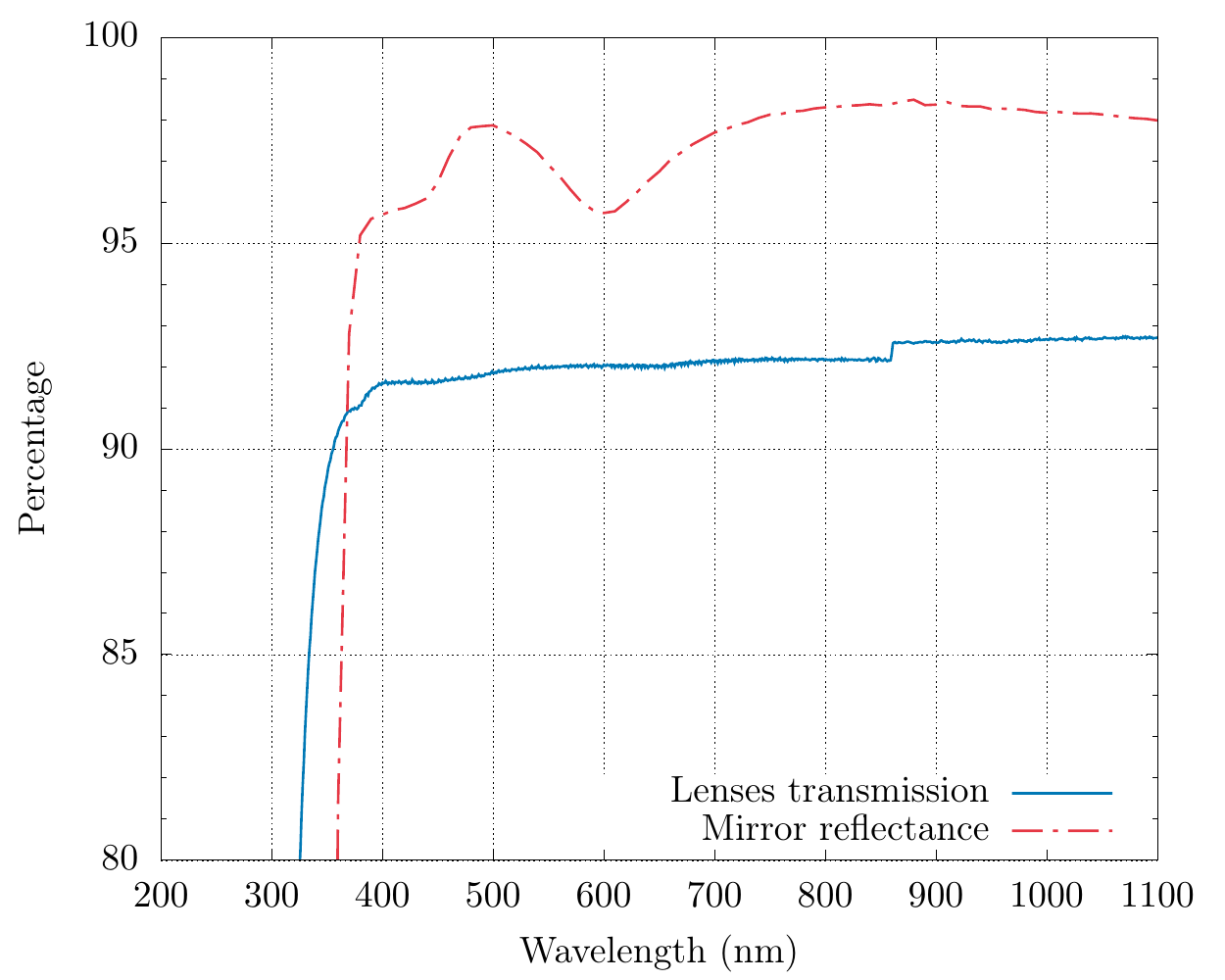}
\caption{ Transmission curve of N-BK7 borosilicate glass (blue solid) and reflectance curve of first surface protected silver coated mirror (red dashed). Data from Thorlabs Inc.}
\label{fig:trans}
\end{figure}

The final design of the light source assembly can be seen in \cref{fig:lightsource}. Lens (L1) is plano-convex with a diameter of $75\,$mm and a focal length of $150\,$mm, and is placed at an equal distance ($150\,$mm) after the secondary focus of the ellipsoidal mirror, almost parallelizing the light beam. Lens (L2), also a plano-convex of $75\,$mm in diameter but with a focal length of $1000\,$mm, is placed 500mm after lens (L1), and acts as a pre-focusing lens, in an effort to prevent beam divergence. Mirror (M1) is a silver coated first surface flat mirror used to divert the beam downwards and its reflectance curve is shown in \cref{fig:trans}. Lens (L3) is a plano-convex lens with a diameter of $150\,$mm and a focal length of $300\,$mm and is placed $600\,$mm from Lens (L2) focusing the light to a  spot of $35\,$mm in diameter onto the test table. All the lenses we used are made of uncoated N-BK7 borosilicate glass, which was chosen as the optimal material for our purpose due to the relatively flat spectral transmission  (\cref{fig:trans}). 

An enclosure was made with aluminum strut profile beams mounted on the lamp housing and aluminum panels. The optical rail onto which the optical elements are mounted is adjustable in three dimensions to allow for precise alignment. Four 22-mm-diameter filtered fans are installed on the bottom side of the structure which, in conjunction with another two high flow fans on the top, provide sufficient ventilation and cooling of the optical and supporting elements.

\subsection{Measuring and monitoring devices}

For the purposes of our experiments the monitoring and sensing capabilities necessary fall under three categories: imaging, temperature sensing and outgassing analysis. 

\subsection*{Imaging}

To achieve a detailed recording of the experiments both in the spatial and time domains we implemented a setup that makes use of four cameras in total, two regular recording cameras and two high-speed cameras. 

The regular cameras of our choice are the Panasonic TZ-100, with 20.1-megapixel sensors. These cameras are compact and lightweight enough to be easily mounted at the top viewing windows on the left and back sides of the chamber. They offer a wide range of zoom capability allowing to frame each experimental run optimally to avoid wasted pixels. To avoid over-exposure due to the strong illumination of the samples, they are fitted with neutral density filters with an optical density of 1.3 (5\% transmission). These cameras are used for the entire duration of each experiment, and apart from capturing the physical alterations of the samples from two different angles, they are also used to record potential changes in the color of the samples.

To monitor faster processes happening during the experiments, we make use of two Chronos 1.4 high-speed cameras made by Krontech Inc. These are 1.4-megapixel monochromatic cameras, with an internal RAM buffer of 32~GB, fitted with Pentax C31204 zoom lenses (12.5--75 mm) and OD 0.6 neutral density filters. The size of the RAM buffer is the limiting factor of the duration of high speed video able to be recorded, which is a function of resolution and frame rate. It is possible though to adjust these two variables to achieve the best quality/duration ratio for each experiment. As an example, at full resolution ($1280\times1024\,\textrm{pix}$) they are capable of recording 16.33~seconds of video at 1069 frames per second (fps). One of these cameras is mounted at the top viewing port of the chamber, while the other is mounted at the bottom viewing port of the left side, offering an edge-on view of the sample. Additionally the cameras are able to broadcast in real time via network and also store the entire experiment at a relatively low frame rate of 60  fps, independently of the high speed recording, which results in four viewing angles of regular video footage in total.

\subsection*{Temperature sensing}

In order to monitor the temperature of the samples during the experiments, SHINeS is equipped with an array of 24 type-K thermocouples, made by Pentronic AB with exposed measuring junctions and glass fiber insulation. They are connected through the D-sub 25 connectors to a data logger which is operated via the vacuum chamber's HMI. The small size of the measuring junctions combined with the number of available thermocouples allows for localised measurements of the internal temperature of the samples and its gradient over the duration of the experiments. 

Additionally, we have a FLIR One PRO thermal imaging camera mounted at the interior of the chamber and connected via the USB 3.0 port, which can be used to monitor the surface temperatures of the samples, but its measuring capability is limited to $T\leq\ang{400}$C.

\subsection*{Outgassing analysis}

One important aspect of the experiments we have scheduled to perform with this system is the outgassing of volatiles from the mineral samples under test. Such a process could be responsible for the fracturing of the samples or even the creation of thrust on small particles, both of which can have a great impact on the evolution of asteroid surfaces. In order to study this, the vacuum chamber is equipped with a HAL-200 residual gas analyser by Hiden Analytical. It has a Faraday cup detector which is capable of operating at pressures that do not exceed $10^{-4}\,$mbar to detect partial pressures down to $5\times10^{-12}\,$mbar in a molecular mass range of 1--200 atomic mass units (amu). The accompanying software allows a wide range of measurement parameters to be adjusted and also performs automatic molecular analysis to determine the composition of the measured gases.
\begin{table}[]
\caption{Summary table of all the main components of SHINeS}
\resizebox{\textwidth}{!}{%
\begin{tabular}{|l|l|l|l|l|}
\hline
\textbf{Component}   & \textbf{Manufacturer} & \textbf{Model}    & \textbf{Description}                                                   & \textbf{Notes}                                                                                                     \\ \hline
Vacuum chamber       & Nanovac AB            & Custom made       & 1m x 1m x 1m steel chamber                                             & \begin{tabular}[c]{@{}l@{}}Adjustable cooled table\\ 6 viewing ports \\ customizable interconnections\end{tabular} \\ \hline
Roughing pump        & Hanbell Co. Ltd.      & PSE-80-NC         & Dry Screw Vacuum Pump                                                  & 1300 L/min pumping speed                                                                                           \\ \hline
High vacuum pump     & Leybold GmbH          & MAG W 1300 iP     & Turbomolecular vacuum pump                                             & \textless{}10\textasciicircum{}-8 mbar ultimate pressure                                                           \\ \hline
Mass spectrometer    & Hiden Analytical      & HAL-200           & Residual gas analyser                                                  & \begin{tabular}[c]{@{}l@{}}Faraday cup\\ 1-200 amu detectable mass range\end{tabular}                              \\ \hline
Lamp housing         & Ernemann GmbH         & Xenosol 7000      & Xenon-arc lamp housing                                                 & \begin{tabular}[c]{@{}l@{}}Up to 7000 W electrical power\\ ellipsoidal IR/UV-cut coated mirror\end{tabular}        \\ \hline
Lamp power supply    & IREM SpA              & EX-170GM3E        & Electronic power supply                                                & \begin{tabular}[c]{@{}l@{}}Up to 7300 W electrical power\\  electronic power control\end{tabular}                  \\ \hline
Optical path         & Built in-house        & Custom made       & Beam formation optics                                                  & \begin{tabular}[c]{@{}l@{}}NBK-7 lenses\\ silver coated first surface mirror\end{tabular}                          \\ \hline
High speed imaging   & Kron Tech. Inc.       & Chronos 1.4       & \begin{tabular}[c]{@{}l@{}}High speed\\ monochrome camera\end{tabular} & \begin{tabular}[c]{@{}l@{}}1.3 megapixel sensor \\ 1280×1024 @ 1057fps \\ 32 GB memory\end{tabular}                \\ \hline
Regular imaging      & Panasonic             & TZ-100            & Portable digital camera                                                & 20.1 megapixel sensor                                                                                              \\ \hline
Temperature sensing  & Pentronic AB          & 6101E-19KK-1000-0 & Type K thermocouple                                                    & Array of 24 thermocouples                                                                                          \\ \hline
Temperature sensing  & Teledyne FLIR         & Flir One Pro      & Portable thermal camera                                                & Up to $400^o\,\mathrm{C}$                                                                                          \\ \hline
\end{tabular}%
}

\end{table}
\section{Calibration and testing}

The first step in the calibration procedure of the solar simulator was to achieve the best possible light-spot characteristics, in terms of light uniformity and power output. 

\subsection*{Uniformity of the light spot}

The main adjustments affecting the uniformity of the light are the positioning of the Xenon lamp in the lamp housing relative to the ideal position of the primary focus of the ellipsoidal mirror (M1), and the distance between the sample plane and the focusing lens (L3). In order to achieve the best possible result for our needs, we resorted to an imaging method which allowed us to analyse the spot intensity profile and make the necessary adjustments. Similar methods have also been used in other designs for solar simulators \citep{thomas2011,Craig2010,SARWAR2014}. 

To this purpose we made use of a piece of aluminum oxide  ($Al_2O_3$) fiber board, which exhibits Lambertian reflectance characteristics, placed on the chamber's table.  A Canon EOS-650D digital camera was placed approximately $60\,$cm from the light spot at a $\ang{45}$ angle, and was used to capture raw images (18 Mpixel, 14-bit Canon original grayscale) of the illuminated target. The main advantage of using raw file format is that the images taken are not gamma-corrected ($\gamma=1$), which means that the value of each individual pixel is directly proportional to the irradiance.

The stored images were then processed with GNU Image Manipulation Program (GIMP) in order to restore the spot image to a circular shape (from elliptical due to the angle of view), crop the useful part of the images and finally downscale them to $200\times200$ pixels, using cubic interpolation of the pixel values, and were exported in TIFF file format. Next we imported the images into Mathematica, where they were handled as three dimensional arrays of (x,y,V) data (x,y coordinates and the pixel values in 8-bit depth), which allowed us to essentially plot the normalised irradiance distribution of the light spot.

\begin{figure}[h]
\centering
\includegraphics[width=0.47\textwidth]{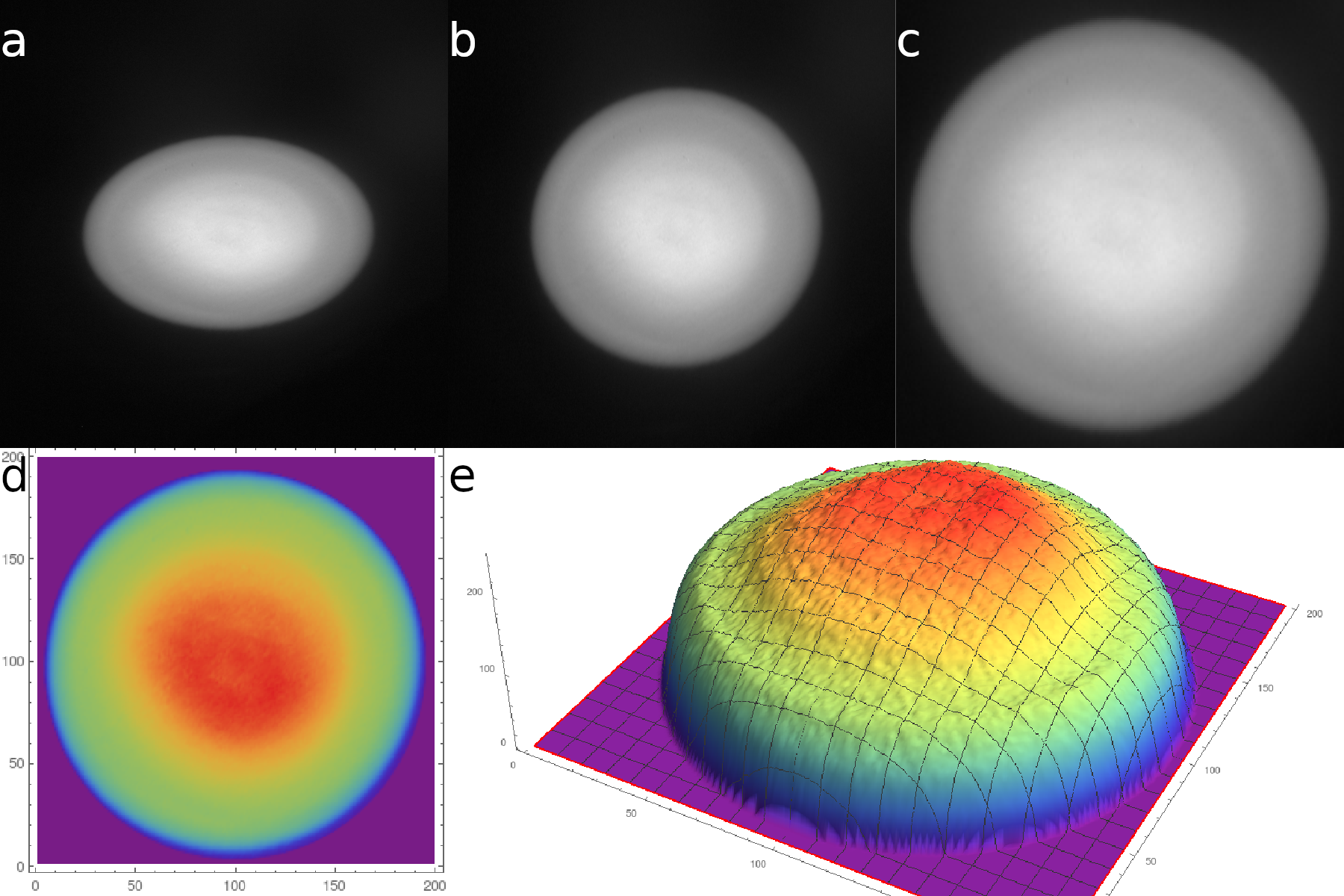}
\caption{Steps of image processing for measuring the uniformity of the light spot: a) Original image, b) Circularized image, c) Cropped and downscaled image, d) Density plot of the luminance, e) 3-D plot of the luminance.}
\label{fig:spotsteps}
\end{figure}

The whole process was repeated while making adjustments to the position of the lamp and the table height until a satisfactory uniformity was achieved. The steps of the process and the final result can be seen in Fig.~\ref{fig:spotsteps}. Although the irradiance is not totally uniform, the distribution is smooth enough with the central, and brightest region of the spot especially homogeneous with a variation in irradiance of about 3.5\%.

\subsection*{Thermal power output}

To measure the optical power output of the system we used a PRONTO-10K thermal power detector manufactured by Gentec-EO Inc., which is a thermopile-type detector with a circular sensing area of $55\,$mm in diameter. The power meter is placed on the working table inside the chamber with its receiving surface at the same position as the uniformity calibration board and the future samples and it is exposed to the incident beam using the lamp housing's manual shutter. It automatically detects the presence of light above some threshold and measures the temperature gradient for a period of 5 seconds, after which it outputs the thermal power of the incident beam taking into account its factory calibration.

The input parameter for these measurements is the DC current supplied to the lamp by the power supply. The power supply sets the operating voltage automatically to maintain a stable current flow. For the lamp we use, the allowed operating range is 108--163~A, yielding an actual electrical power consumption ranging from 3456~W to 6520~W. We performed five measurements of the thermal power delivered to the detector in this current range in increments of 5~A. The current can be controlled in smaller increments by the power supply, but we considered 5~A increments to be enough for our purposes here.

\begin{figure}[h]
\centering
\includegraphics[width=0.47\textwidth]{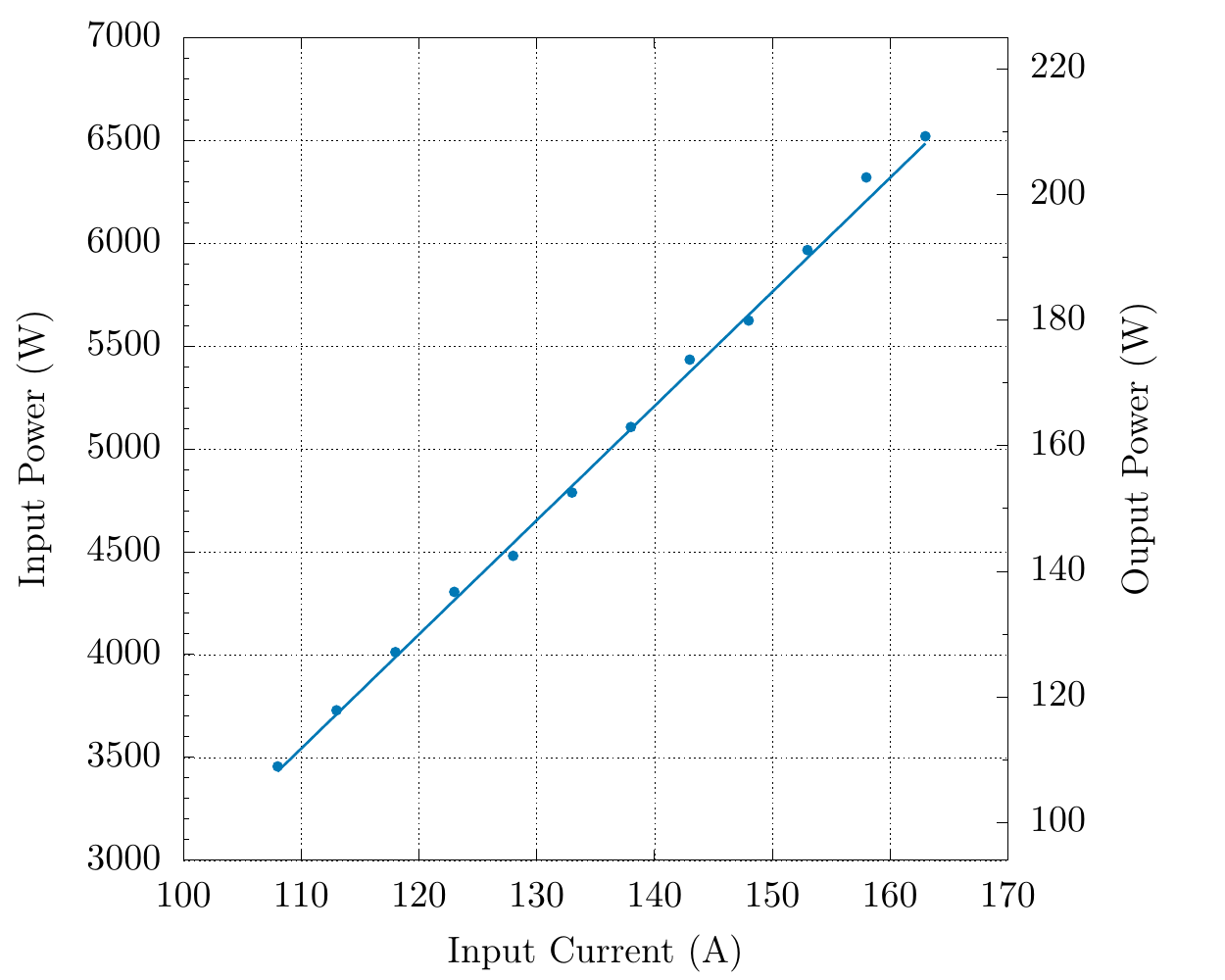}
\caption{Power transfer characteristics of the solar simulator.}
\label{fig:power}
\end{figure}

The results are plotted as measured thermal power versus input current and electrical power (\cref{fig:power}). Unfortunately, due to the fact that the power detector is rated for 10000~W of total power, its displayed measurement accuracy is 1~W which resulted in identical measurements in each of the five instances at any input current value. This fact did not allow for a proper error analysis, but it also means that any variations of the output optical power at any level are smaller than 1~W.  The average efficiency of the system was calculated to be approximately 3.17\%. Although it seems quite low, it is to be expected given the known low efficiency of Xenon arc lamps, the fraction of light lost in the lamp housing due to the geometry of the ellipsoidal mirror (about 38\%), and losses in the rest of the lightpath.  

Having in mind that the irradiance distribution is not uniform across the whole area of the light spot, the average irradiance, which we can compute by simply dividing the measured optical power with the area of the light spot, is not a very useful quantity, since we are mostly interested in the central region of the light. More suitable quantities for the characterisation of the system and further applications are the peak irradiance and the average irradiance of only the central part of the light spot. These have been computed by combining the  measurements of the total optical power delivered to the target at maximum input electrical power with the light-spot-uniformity analysis presented above.

Since each pixel value (V) is proportional to the irradiance at its location, the total optical power can be associated with the volume under the surface defined by the (x,y,V) coordinates, which can be calculated by finding an interpolating function and then integrating over the area. A simpler way is to use directly the discrete array of pixels. If we consider the area represented by each pixel equal to 1, the volume under the surface is simply the sum of all the pixel values in arbitrary units. From the images and the measured actual size of the light spot, assuming a perfectly circular shape, we can compute the pixel density  in pixels per square centimeter: ($D_p=3104.08\,\mathrm{pix/cm^2}$). The peak irradiance can then be calculated as:
\begin{equation*}
    E_{peak,max}=\frac{V_{peak}}{\sum_{n=1}^{N_{tot}} V}\times D_p\times \Phi_{tot,max}=37.46\, \mathrm{W/cm^2}\,,
\end{equation*}
where $V_{peak}$ is the maximal pixel value in the image, $\sum V$ is the sum of all pixel values over all the pixels $N_{tot}$ and $\Phi_{tot}=210\,\mathrm{W}$ is the measured optical power delivered across the whole spot at maximum input electrical power. 

We then defined the central region of the spot as the circle with a diameter of 70 pixels, roughly equal to one third of the whole spot. We masked this region from the same image used before in GIMP, and then calculated the optical power delivered to this region as:
\begin{equation*}
    \Phi_{cen,max}=\frac{\sum_{n=1}^{N_{cen}} V}{\sum_{n=1}^{N_{tot}} V}\times\Phi_{tot,max}=43.53\, \mathrm{W}\,,
\end{equation*}
where $N_{cen}$ is the number of pixels of the central region. The mean irradiance in the central region at maximum input electrical power can finally be calculated as:
\begin{equation*}
    E_{cen,max}=\frac{\Phi_{cen,max}\times D_p}{N_{cen}}=35.47\,\mathrm{W/cm^2}
\end{equation*}
This value is equivalent to about 260 times the solar constant ($\mathrm{G_{sc}}$), or a distance of about $0.0625\,\mathrm{au}$ from the Sun. It is trivial to compute the irradiance at different power levels by simply scaling the maximal value, or inversely to compute the necessary input electrical power to achieve a target irradiance. Just for reference we note here that the minimum irradiance achieved by our system in its current configuration, i.e. only by adjusting the electrical power of the lamp at the lowest setting of its operational range, is $E_{cen,min}=18.41\,\mathrm{W/cm^2}\approx 135\,\mathrm{G_{sc}}$, equivalent to a distance of $0.0866\,\mathrm{au}$ from the Sun. Due to the proportionality of the solar irradiance with the inverse of the square of the distance, for such small distances from the Sun a range of about factor 2 in operating  electrical power, translates to a very small range in simulated distance. It is however possible to achieve lower optical power levels, and therefore simulate larger distances, either by attenuating the light with filters, or adjusting the optics to increase the size of the light spot, or by using a lamp with a lower power rating. 

\subsection*{Spectral characteristics}

The final aspect of characterization for our solar simulator is the spectrum of the output light, as we need this to be as close as possible to the extraterrestrial solar spectrum. A known characteristic of the emission spectrum of Xenon-arc lamps is the presence of strong spikes in the infrared, at wavelengths in the range $800--1000\,\mathrm{nm}$, which produce the key difference compared to the solar spectrum. However, the reflective coating of the ellipsoidal mirror in our lamp housing has very low reflectivity in the infrared wavelengths by design, as in its original purpose, the excess infrared  part of the light would only cause problems to the film projection.

We made use of an Ocean Optics HR4000 spectrometer to measure the spectrum of the solar simulator and verify that it is suitable for our applications. We took measurements of the spectra of the light at two points, directly from the Xenon-arc lamp, and at the final light spot location inside the vacuum chamber using an optical fiber to directly collect incident light. We also took a measurement of direct sunlight to have a qualitative reference. Since our only intent was to compare the three spectra measured by the same device, there was no need to use a reference light source and calibrate the instrument as one would do for absolute irradiance measurements. Therefore, we could just use the direct readings in photon counts which, divided by the wavelength, give values proportional to the irradiance.

\begin{figure}[h]
\centering
\includegraphics[width=0.47\textwidth]{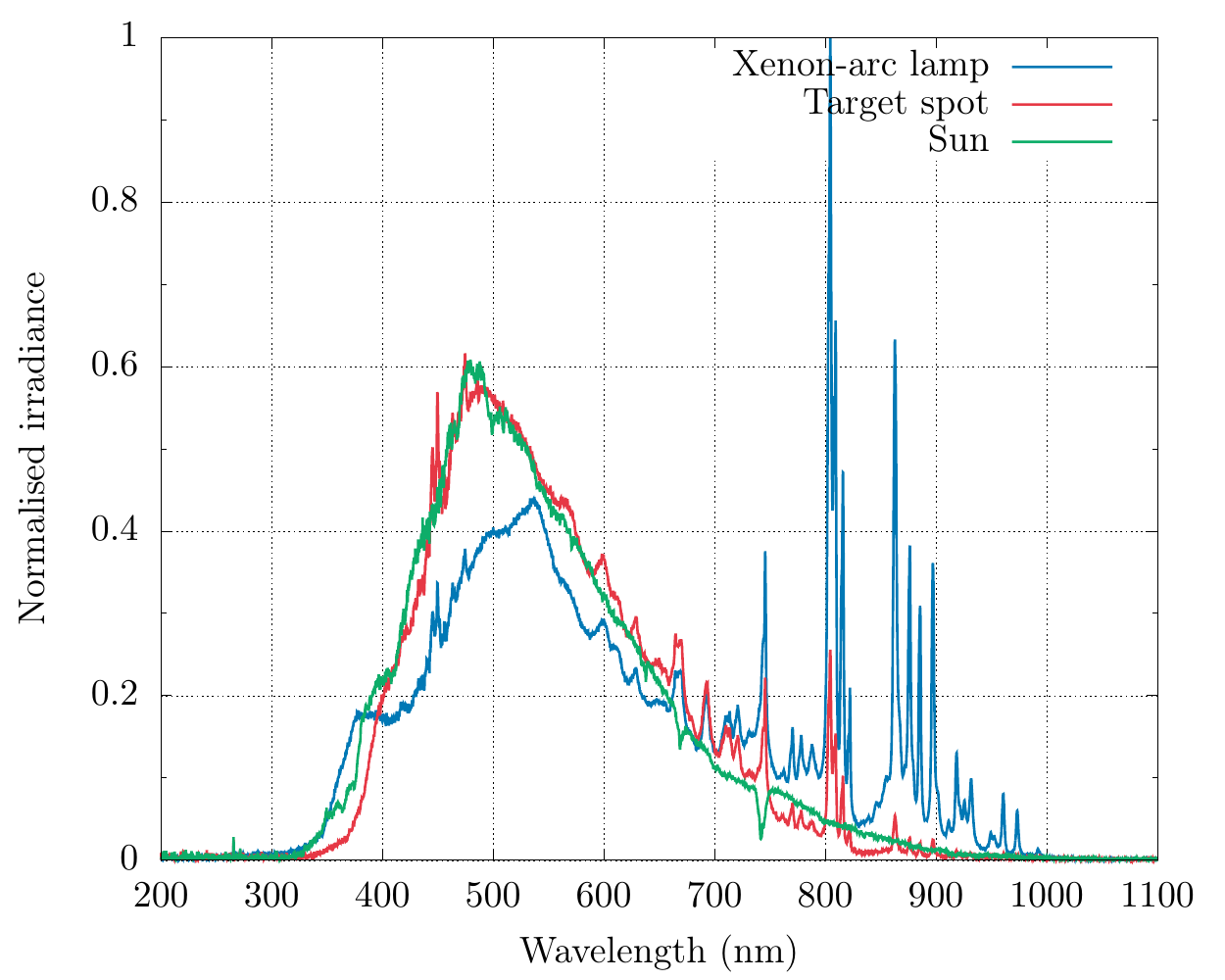}
\caption{Comparison of the measured spectra, normalised on output optical power. The blue line represents the spectrum taken directly from the lamp, the red line represents the output spectrum and the green line the solar spectrum.}
\label{fig:spectra}
\end{figure}

It is evident that the infrared spikes of the lamp's spectrum are indeed attenuated to a great degree, and the resulting spectrum resembles the solar one to a very satisfactory degree, for all practical purposes (\cref{fig:spectra}).

\section{Testing}

Finally, we performed a test experiment to verify that all the sub-systems of SHINeS perform as expected. To this purpose we used a small piece of Serpentine, a hydrous phyllosilicate mineral with a generalized formula: $(\mathrm{Mg,Fe,Ni,Mn,Zn)}_{2-3}\mathrm{(Si,Al,Fe)}_2\mathrm{O}_5\mathrm{(OH)}_4$ which forms on Earth by hydrothermal metamorphism of ultramafic rocks. The serpentine group is also the most common among the hydrated minerals found on meteorites \citep{Rubin1997}, therefore it is believed to also be part of the composition of C-type asteroids.

The sample was weighed before being placed into a sample holder made from aluminum oxide fiber board, at $14.63\,\mathrm{g}$. A small hole was drilled at the underside of the sample to allow for one of the thermocouples to be placed approximately at the center of the sample, in order to measure the internal temperature. The following day, after approximately 21 hours of pumpdown operation the vacuum chamber had reached a pressure of $1.02\times10^{-5}\,\mathrm{mbar}$ ,which was a good point for starting the test. The light source was set at its lowest setting and the sample was irradiated at this level for about 10 minutes, after which we increased the optical power of the light source to its maximum level and left it running for another 20 minutes. 

\begin{figure}[h]
\centering
\includegraphics[width=0.47\textwidth]{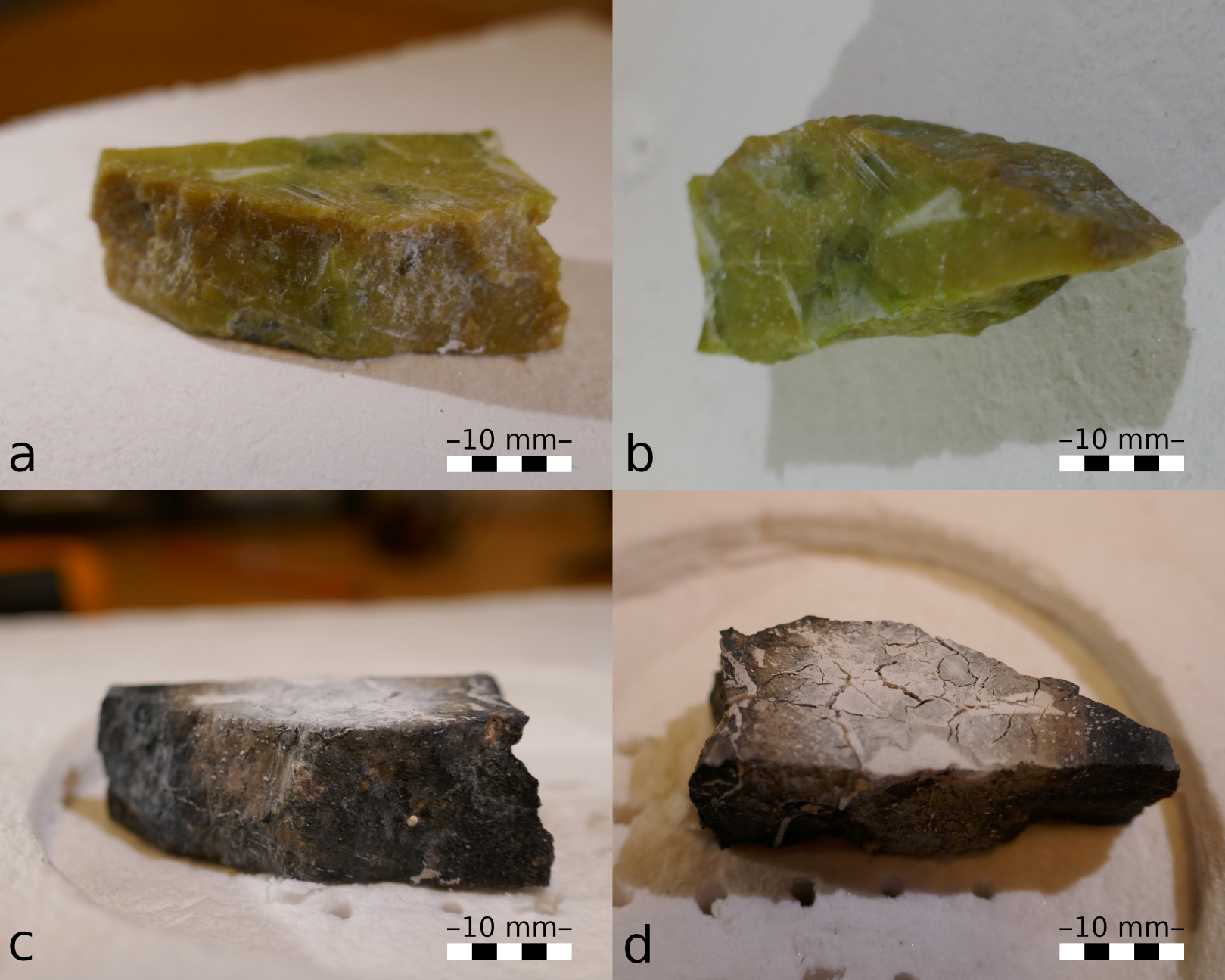}
\caption{Side(left) and top (right) views of the Serpentine sample before (a,b) and after the experiment (c,d).}
\label{fig:sample}
\end{figure}

The serpentine sample is shown in \cref{fig:sample} before and after the testing experiment, in two different views. The change in color of the whole sample is immediately noticeable, changing from green overall to dark gray/brown. Furthermore, its top surface, which was illuminated, and especially the part which was exposed to the highest optical power flux (\cref{fig:spotsteps}), has turned almost white. The layer of white material forming on the surface increases the albedo of the sample. If this is a typical phenomenon on the surfaces of low-albedo, carbon-rich asteroids that have orbits with small perihelion distances, it could provide an alternative explanation for the apparent lack of low-albedo asteroids on orbits with small perihelion distances \citep{Granvik2016}: low-albedo asteroids have not been destroyed but just turned into high-albedo asteroids. However, more extensive laboratory studies are required to properly test the hypothesis put forward. In the top part of the sample we also witness the formation of cracks with a pattern compatible with desiccation cracking \citep{Jewitt2013}. 

\begin{figure}[h]
\centering
\includegraphics[width=0.5\textwidth]{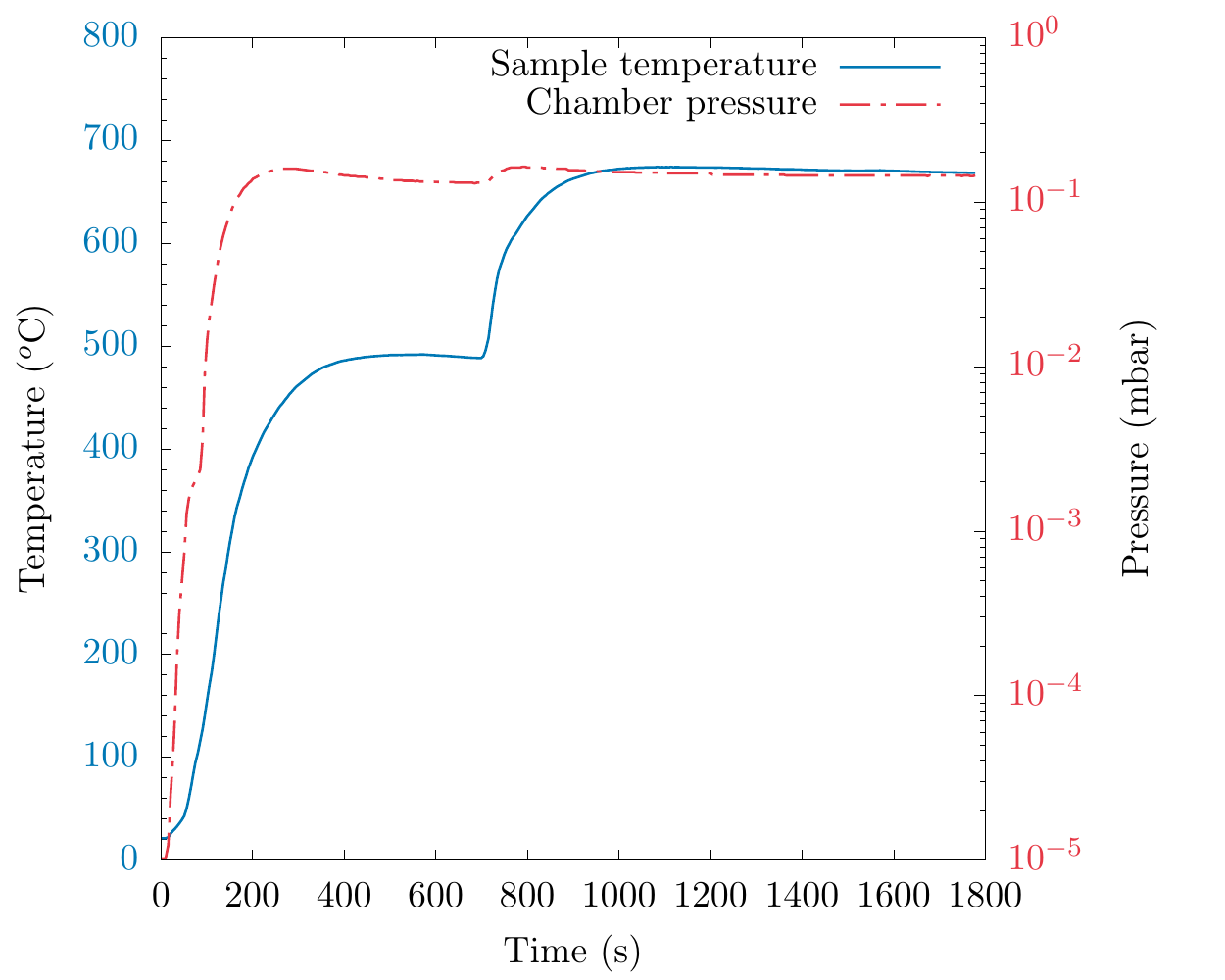}
\caption{The internal temperature of the sample (blue solid) and chamber pressure (red dashed) during the test experiment.}
\label{fig:test}
\end{figure}

The internal temperature of the sample as well as the chamber pressure are shown in \cref{fig:test}. We see that the sample reached a temperature of $490^{\mathrm{o}}\mathrm{C}$ after about 8 minutes of exposure to the low  optical power setting, and $675^{\mathrm{o}}\mathrm{C}$ after  about 6 minutes at the high power setting. We also notice the rapid increase of the chamber pressure, very quickly after the beginning of the exposure, despite the continuous operation of the pumps.  This happens due to the high volatile content of the Serpentine sample, which outgasses significantly under heating, despite its rather small mass.

\begin{figure*}[h]
\centering
\includegraphics[width=\textwidth]{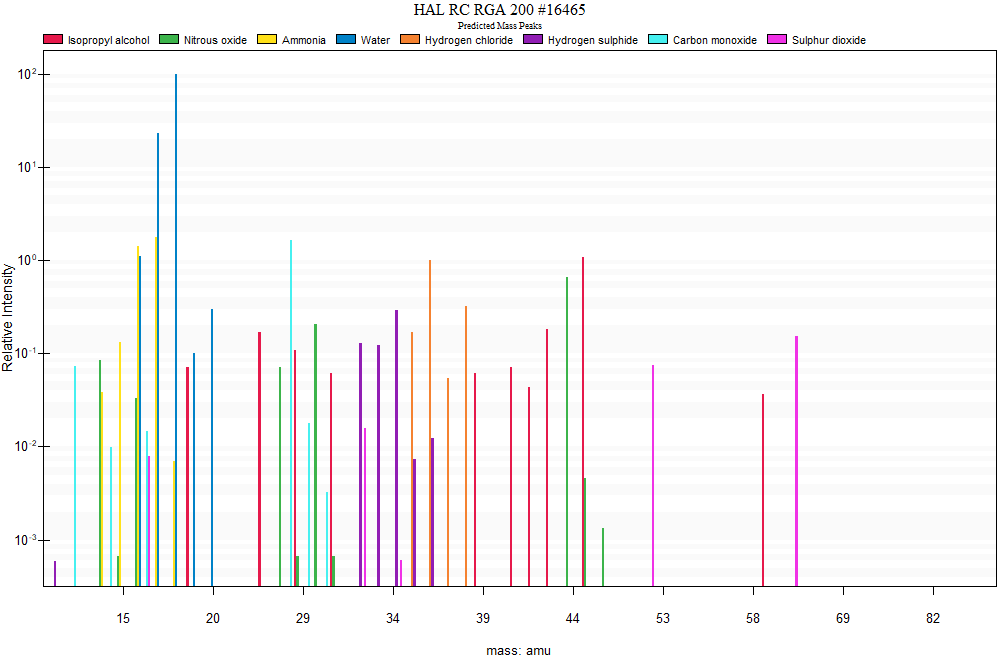}
\caption{Example of the residual gas analysis results. Note that only the most abundant components are plotted here to maintain readability.}
\label{fig:rga}
\end{figure*}

A simple representation of the RGA results is presented in \cref{fig:rga}, showing the relative height of the peaks corresponding to each of the components. The acquisition of the measurements was performed soon after the illumination of the sample began, with 10 scans per amu. The relative composition is afterwards calculated automatically by the software and it is also corrected for the relative sensitivity of the instrument to each component, shown in \cref{table:rga}. The most abundant component is water, which is expected due to both the predicted outgassing from the sample, but also because water vapor is the major residual air component in all vacuum systems. The sample weight after the experiment was measured to be $12.38\,\mathrm{g}$, losing $18.17\%$ of its mass, which by the RGA result can safely be attributed to loss of water content. Many of the other components are also originating from the system itself and not the sample. These are also expected in all vacuum systems, either being contaminants from the manufacturing process or other substances such as lubricants or cleaning agents. For example, isopropyl alcohol was used for cleaning the table and chamber door seal. Therefore the data shown in \cref{fig:rga} and \cref{table:rga} are given only as an example of the capabilities of the RGA and the automatic analysis of its accompanying software. In order to get more useful quantitative results regarding the outgassing, residual gas analysis can be performed before initiating each of the experiments to obtain a base reading of the  chamber's state, and then compare it to the data acquired during the illumination phase.

\begin{table}[]
\caption{Composition results of the residual gas analysis}
\label{table:rga}
\centering
\begin{tabular}{|l|l|l|l|}
\hline
\textbf{Component} & \textbf{Formula} & \textbf{\%} & \textbf{Corrected \%} \\ \hline
Water              & $\mathrm{H_2O}$           & 93.82       & 95.51                 \\ \hline
Ammonia            & $\mathrm{NH_3}$           & 1.64        & 1.16                  \\ \hline
Carbon monoxide    & $\mathrm{CO}$             & 1.54        & 1.34                  \\ \hline
Isopropyl alcohol  & $\mathrm{C_3H_8O}$        & 1.02        & 0.93                  \\ \hline
Hydrogen chloride  & $\mathrm{HCL}$            & 0.94        & 0.54                  \\ \hline
Nitrous oxide      & $\mathrm{N_2O}$           & 0.62        & 0.34                  \\ \hline
Hydrogen sulfide   & $\mathrm{H_2S}$           & 0.28        & 0.11                  \\ \hline
Sulfur dioxide     & $\mathrm{SO_2}$           & 0.14        & 0.06                  \\ \hline
\end{tabular}
\end{table}

\section{Conclusions}
We have presented the development of SHINeS, a complete system to emulate the thermal environment in the immediate neighbourhood of the Sun, and the measurement and monitoring capabilities necessary for the planned experiments to better understand the destruction mechanisms affecting NEAs on orbits with small perihelion distances.

We have tested and demonstrated that the solar simulator can achieve a maximum irradiance level corresponding to a distance of $0.0625\,\mathrm{au}$ from the Sun, in a uniform enough manner to allow for meaningful interpretation of future experiment results. The spectrum of the light closely matches that of the Sun. The characteristic emission peaks of the Xenon-arc lamp are attenuated well enough by the coating of the ellipsoidal mirror, resulting on a good approximation of the AM0 spectrum for all practical applications. 

We have verified by performing a test experiment that all monitoring sub-systems work as intended, and the preliminary results regarding the effects of intense solar radiation on asteroid-like materials are more or less in accordance with expectations.

One drawback of SHINeS is the inability to maintain high vacuum when substantial amounts of volatiles are evaporated due to the high heat, as we witnessed that the pressure rises by four orders of magnitude very rapidly. However, the sample used (Serpentine) is by far the richest in water content among the minerals planned to be used in future experiments, and its size was rather large (larger than the light spot itself and substantially thick), so this can be considered as a marginal experiment. The issue can be rectified in future experiments by using smaller samples and by adjusting the algorithm controlling the pumping procedure. Ultimately, the system design is flexible enough, so that it is even possible to explore other options for the pumping sub-systems, such as installing larger pumps or pumps using different technology, in order to increase the total pumping capacity if it would be necessary for our experiments. 

\section*{Acknowledgements}

Funding for the experiment setup was provided by the Kempe Foundation and the authors are grateful for the flexibility  expressed by the Kempe Foundation during the COVID-19 pandemic. All authors were funded by the Knut and Alice Wallenberg foundation. MG also acknowledges funding from the Academy of Finland. The authors are grateful for the expert advice provided by C.~Dreyer, and to the two anonymous reviewers for their constructive comments and suggestions.
\def\newblock{\ }
\bibliographystyle{jphysicsB}
\bibliography{bib.bib}
\end{document}